\documentclass[smallcondensed]{svjour3}

\usepackage{texdraw}
\usepackage{graphicx}
\usepackage{dcolumn}
\usepackage{bm}
\usepackage{amsmath}
\usepackage{amssymb}
\usepackage{subfigure}
\usepackage{url}
\usepackage{multirow}

\begin{document}

\title{Permeability Description by Characteristic Length, Tortuosity, Constriction and Porosity}
\titlerunning{Permeability Description}

\author{Carl Fredrik Berg}
\authorrunning{C.\ F.\ Berg}
\institute{Carl Fredrik Berg \\ Statoil R\&D Center, Arkitekt Ebbels veg 10, Rotvoll, 7053 Trondheim, Norway \\ cfbe@statoil.com, carlpaatur@hotmail.com}

\date{Accepted: 14 March 2014 in Transport in Porous Media}

\maketitle

\begin{abstract}
In this article we investigate the permeability of a porous medium as given in Darcy's law. 
The permeability is described by an effective hydraulic pore radius in the porous medium, the fluctuation in local hydraulic pore radii, the length of streamlines, and the fractional volume conducting flow. The effective hydraulic pore radius is related to a characteristic hydraulic length, the fluctuation in local hydraulic radii is related to a constriction factor, the length of streamlines is characterized by a tortuosity, and the fractional volume conducting flow from inlet to outlet is described by an effective porosity. The characteristic length, the constriction factor, the tortuosity and the effective porosity are thus intrinsic descriptors of the pore structure relative to direction. 
We show that the combined effect of our pore structure description fully describes the permeability of a porous medium. The theory is applied to idealized porous media, where it reproduces Darcy's law for fluid flow derived from the Hagen-Poiseuille equation. We also apply this theory to full network models of Fontainebleau sandstone, where we show how the pore structure and permeability correlate with porosity for such natural porous media. This work establishes how the permeability can be related to porosity, in the sense of Kozeny-Carman, through fundamental and well-defined pore structure parameters: characteristic length, constriction, and tortuosity.

\end{abstract}

\section{Introduction}\label{sec:introduction}

Understanding the process of flow through porous media is of great importance in many fields, including petroleum engineering and hydrology. 
Slow viscous fluid flow in porous media is traditionally described by Darcy's law, a proportional relation that links the fluid discharge $Q$ to an applied piezometric head (hydraulic head) difference $\Delta h$:
\begin{equation}
k= -\frac{Q \mu \Delta s}{A \rho g \Delta h},
\label{eq:darcy}
\end{equation}
where $A$ is the cross-sectional area of the porous medium, $\Delta s$ is the length of the porous medium in direction of applied head difference, $\mu$ is the constant fluid viscosity, $\rho$ is the constant fluid density, $g$ is acceleration due to gravity, and $k$ is a constant for the porous medium called (intrinsic) \emph{permeability} \cite{darcy1856determination,bear1988dynamics,dullien1992porous}. For simplicity, $\rho g h$ will also be called the piezometric head throughout this article.

A fundamental question in flow through porous media is how the permeability $k$ can be related to the porosity $\phi$ through well-defined parameters of the pore structure. A well-known relation of this kind is the semi-empirical Kozeny-Carman equation. It was first proposed by Kozeny \cite{kozeny1927ueber} as
\begin{equation}
k \propto \tau \frac{\phi^3}{(1-\phi)^2}d_w^2,
\label{eq:kozeny}
\end{equation}
where $d_w$ is an effective grain size, and $\tau$ is a tortuosity of the porous medium describing the relative difference between the microscopic (interstitial) head gradient along the streamline and the macroscopic head gradient. Kozeny derived his equation assuming that the porous medium could be viewed as a bundle of streamtubes \cite{kozeny1927ueber}.

Carman noted that linking the microscopic fluid velocity to the Darcy velocity for the porous medium involves scaling with the factor $\tau$ \cite{carman1937fluid}. Carman therefore modified Kozeny's Eq.\ \eqref{eq:kozeny} by multiplying with the tortuosity $\tau$ \cite{carman1937fluid}:
\begin{equation}
k \propto \tau^2 \frac{\phi^3}{(1-\phi)^2}d_w^2.
\label{eq:carman}
\end{equation}
For a monodisperse sphere pack we have $d_w = 6(1-\phi)/S_0$, where $S_0$ is the specific surface area \cite{kozeny1927ueber}. This leads to a more general form of the Kozeny-Carman equation:
\begin{equation}
k = c_0 \tau^2 \frac{\phi^3}{S_0^2} = c_0 \tau^2  r_h^2 \phi,
\label{eq:kozeny-carman}
\end{equation}
here $r_h = \phi/S_0$ is the (mean) \emph{hydraulic radius} and $c_o$ is a coefficient called \emph{Kozeny's constant} \cite{kozeny1927ueber,carman1937fluid,bear1988dynamics,dullien1992porous}.

The hydraulic radius $r_h$ is assumed to represent an effective pore radius of the porous medium. It is a purely geometric length which does not take into account the effect on permeability from pore size variation or connectivity. Others have proposed length-scales more suitable for permeability description, among these the smallest pore along the most conductive percolating pathway \cite{katz1986quantitative},  nuclear magnetic resonance relaxation time \cite{banavar1987magnetic}, or grain size distribution \cite{berg1970method}. 
Johnson \emph{et al.}\ \cite{johnson1986new} suggested a length by weighting with the electric field, thereby related to (electrical) transport and thus a \emph{dynamical} length in contrast to the \emph{geometrical} hydraulic radius $r_h$. 
This length has been shown to be a better permeability descriptor than the hydraulic radius $r_h$ \cite{schwartz1993cross}, however, there is no fixed relation between electrical conductance and fluid flow \cite{saeger1991flow}. A dynamical length linked to fluid flow instead of electrical conductance was introduced in \cite{bear1967generalized}. This length was derived from the microscopic hydraulic conductance \cite{bear1967generalized,bear1988dynamics}, thus descriptive of fluid flow in porous media.

For a porous medium, the (hydraulic) tortuosity is a measure of the microscopic flows deviation from the direction of the applied piezometric head difference, reflected by the length of the microscopic streamlines \cite{bear1988dynamics,dullien1992porous,adler1992porous}. Tortuosity has been defined as $\Delta s / s_e$, where $\Delta s$ is the length of the porous medium in the direction of applied piezometric head difference, and $s_e$ is the effective streamline length \cite{bear1988dynamics,koponen1996tortuous}. There is ambiguity associated with the derivation of $s_e$ \cite{bear1988dynamics,koponen1996tortuous}, however most formulations are a weighted average of the streamline lengths \cite{duda2011hydraulic}. 

Furthermore, the constricting and expanding nature of pore channels converges and diverges the streamlines, which leads to variation in fluid velocity along the streamlines and decrease the permeability. This effect has been treated for simplified porous media \cite{dullien1992porous}, for more complex material the standard deviation of the cross-sectional area has been used as an estimate \cite{schopper1966theoretical}, while for electrical conductance the effect of pore channel variation has been described in general \cite{berg2012}. We follow Ref.\ \cite{dullien1992porous} and use the term \emph{constriction factor} to account for this porous medium property. Note that the term \emph{constrictibility} is common when considering effective diffusion in porous media, see e.g.\ Ref.\ \cite{van1974analysis}. The effect on transport from constrictions is frequently lumped together with the effect from tortuosity, see e.g.\ Refs.\ \cite{schwartz1993cross,bear1967generalized,bear1988dynamics}. However, the pore structure effect on permeability from these two distinct geometrical properties can be separated for any porous medium, as shown in this article.

The (geometrical) porosity $\phi=\Omega/V$ is the fraction of pore space $\Omega$ in the porous medium of total volume $V$. The pore space $\Omega$ is sometimes interpreted as the connected pore space, thus both the permeability and porosity vanish at a percolation threshold at a finite geometrical porosity \cite{mavko1997effect}. Other authors have also excluded dead-end pores \cite{bear1988dynamics,guo2012dependency}, or considered an effective porosity based on the streamlines connected both to the inlet and the outlet of the porous medium \cite{duda2011hydraulic,koponen1997permeability}.

A large body of literature exists on the relation between macroscopic transport properties and pore structure; the interested reader is referred to \cite{dullien1992porous} and \cite{bear1988dynamics} for reviews of numerous relations. The recent progress in computing and imaging provides tools for advances on the topic. A number of methods for statistically \cite{adler1990flow,yeong1998reconstructing,jiao2009superior} or process based \cite{oren2002process} reconstruction of three-dimensional porous media from two-dimensional thin section images has been proposed. Advances in micro-computed tomography (mCT) have made it possible to directly image many types of natural porous media with sufficient resolution to represent the three-dimensional pore structure \cite{schwartz1994transport,arns2001accurate}. For flow and transport simulations network analogs have been widely used to represent the pore space \cite{fatt1956network,blunt2001flow,oren1998extending}, while the full microscopic pressure and velocity fields can also be computed for both idealized models \cite{lemaitre1990fractal,schwartz1993cross,zhang1995direct} and for natural porous media such as reservoir rocks \cite{arns2001accurate,mostaghimi2013computations}.

The aim of this work is to derive a comprehensive relation for porous media between the permeability and porosity from detailed pore structure information. In contrast with existing relations the permeability will be fully defined by separable descriptors of the pore structure without introducing free parameters or constants.
 In Sect.\ \ref{section:local_perm} we introduce a microscopic permeability factor which describes the local contribution to the effectiveness of the pore space to conduct fluid flow. This permeability factor is decomposed into streamlines of the fluid flow, and further factorized into distinct contributions from characteristic length, constriction and tortuosity in Sect.\ \ref{sec:perm_streamline}. In Sect.\ \ref{sec:perm_description} we integrate the characteristic length, constriction and tortuosity from individual streamlines into effective parameters, all pore structure descriptors. In Sect.\ \ref{sec:hagen-poiseuille} we use the Hagen-Poiseuille equation to demonstrate our approach on idealized porous media, while the methodology is applied to network analogs of Fontainebleau sandstone data in Sect.\ \ref{section:bentheimer_network}.

\section{Microscopic (Interstitial) Permeability}\label{section:local_perm}

Consider a porous medium $V$ of length $\Delta s$ in direction of an applied piezometric head difference $\Delta h$, consisting of matrix and pore space $\Omega \subset V$ filled with an incompressible fluid. 
 At the microscopic (interstitial) scale, a slow (creeping) flow is governed by the Stokes equation supplemented by the continuity equation:
\begin{subequations}
\begin{align}
\mu \nabla^2 \vec{u} &= \rho g \nabla h, \label{eq:stokesa} \\
\nabla \cdot \vec{u} &= 0, \label{eq:stokesb}
\end{align}
\label{eq:stokes}
\end{subequations}
where $\vec{u}$ is the microscopic fluid velocity, and $h$ is the microscopic piezometric head \cite{dullien1992porous,bear1988dynamics,adler1992porous}. 
In the following we will refer to Eqs.\ \eqref{eq:stokes} simply as the Stokes equations.
Throughout this article the fluid flow is assumed to be governed by the Stokes equations. There are no time derivative terms in the Stokes equations; hence a constant piezometric head difference $\Delta h$ implies a steady-state flow.

Let $\mathbb{S}$ denote the set of all streamlines $\mathcal{S}$ connected both to the inlet and the outlet of the porous medium, and let $\Omega_s = \{ x \in \mathcal{S} \in \mathbb{S} \}$ be the subset of $\Omega$ where the fluid flows from inlet to outlet. The \emph{effective porosity} is $\phi_s = \Omega_s/V$ \cite{koponen1997permeability,duda2011hydraulic}. Due to the linearity of the Stokes equations, the streamlines $\mathcal{S}$ are independent of the magnitude of applied piezometric head drop $\Delta h$ and the constants $\rho$ and $\mu$, thus $\Omega_s$ and $\phi_s$ are only dependent on pore structure and direction of the applied piezometric head drop. Note that for dead end pores we might have pore space that is not in $\Omega_s$, still the fluid velocity might be non-zero, called reentrant flow in Ref.\ \cite{duda2011hydraulic}.

In our system the piezometric head difference $-\rho g \Delta h$ is the potential that drives the fluid through the porous medium, hence the rate of applied energy is $-\rho g \Delta h Q$. The potential driving the microscopic fluid flow is the piezometric head $\rho g h$, and the rate of work done by this piezometric head potential is given by $- \rho g \nabla h \cdot \vec{u}$.

Applying the divergence theorem, and invoking that the fluid velocity $\vec{u}$ is a solenoidal vector field from the continuity equation Eq.\ \eqref{eq:stokesb}, we obtain
\begin{align}
\int_{\Omega_s} \rho g \nabla h \cdot \vec{u} dV &= \int_{\Omega_s} \rho g \nabla h \cdot \vec{u} + \rho g h(\nabla \cdot \vec{u}) dV \notag \\
&= \int_{\Omega_s} \nabla \cdot \left( \rho g h \vec{u} \right) dV 
= \int_{\delta \Omega_s} \rho g h \vec{u} \cdot \vec{n} dS \notag \\
&= \rho g (h_{out}Q - h_{in}Q) 
= \rho g \Delta h Q,
\end{align}
where $\vec{n}$ is the outward pointing unit normal field of the boundary $\delta \Omega_s$, $h_{in}$ is the piezometric head at inlet, and $h_{out}$ the head at outlet. Here we use that $\vec{u} \cdot \vec{n} = 0$ except at inlet and outlet.

The permeability as given by Darcy's law in Eq.\ \eqref{eq:darcy} can then be expressed as follows:
\begin{equation}
k= -\frac{Q \mu \Delta s}{A \rho g \Delta h} = \phi_s \frac{1}{\Omega_s} \int_{\Omega_s} -\mu \rho g \nabla h \cdot \vec{u} \left( \frac{\Delta s}{\rho g \Delta h} \right)^2 dV.
\label{eq:darcy_frac_to_power}
\end{equation}
We will denote the integrand in Eq.\ \eqref{eq:darcy_frac_to_power} as the \emph{microscopic permeability factor}: 
\begin{equation}
\kappa =  -\mu \rho g \nabla h \cdot \vec{u} \left( \frac{\Delta s}{\rho g \Delta h} \right)^2.
\label{eq:kappa_def}
\end{equation}
The microscopic permeability factor $\kappa$ is then the rate of work done by the piezometric head potential $-\rho g \nabla h \cdot \vec{u}$ multiplied by the factor $\mu \Delta s^2 /(\rho g \Delta h)^2$, however it is a constant only dependent on the pore space $\Omega \subset V$ and the direction of the applied head difference $\Delta h$, in contrast to $-\rho g \nabla h \cdot \vec{u}$.

We derive the \emph{effective permeability factor} by integrating $\kappa$ over $\Omega_s$:
\begin{equation}
\kappa_s = \frac{1}{\Omega_s}\int_{\Omega_s} \kappa dV = -\frac{1}{\Omega_s}\mu \rho g \Delta h Q\left(\frac{\Delta s}{\rho g \Delta h}\right)^2.
\label{eq:kappa_glob_def}
\end{equation}
Since the microscopic permeability factors $\kappa$ are constants only dependent on the pore space and the direction of the applied head difference $\Delta h$, so is the macroscopic permeability factor $\kappa_s$. Moreover
\begin{equation}
k= \kappa_s \phi_s.
\label{eq:darcy_global}
\end{equation}
We have thereby divided the permeability into two factors; the effective porosity $\phi_s$ yielding the pore space fraction where fluid flow from inlet to outlet, and the permeability factor $\kappa_s$ yielding the effectiveness of the pore space $\Omega_s$ to conduct fluid flow. Both $\phi_s$ and $\kappa_s$ are dependent on direction, leading to the anisotropy of the permeability.

Following the same arguments as above we also have
\begin{equation}
k=\hat{\kappa} \phi,
\end{equation}
where
\begin{equation}
\hat{\kappa} = \frac{1}{\Omega}\int_{\Omega} \kappa dV.
\end{equation}
This implies that
\begin{equation}
\int_{\Omega \setminus \Omega_s} \kappa dV = 0,
\end{equation}
even though the permeability factor $\kappa$ might be non-zero in $\Omega \setminus \Omega_s$. Hence reentrant flow does not contribute to the effective permeability factor. In the following we work with streamlines connected to inlet and outlet, thereby excluding the part of the pore space containing reentrant flow in addition to stagnant parts.

\section{Streamline Decomposition}
\label{sec:perm_streamline}

In this section, we show how the permeability factor can be decomposed onto streamlines, and furthermore how the permeability factor for each streamline can be segmented into parts describing the hydraulic conductance, the constrictions and the tortuosity along this streamline.

We can discretize the space $\Omega_s$ into a disjoint union of simply connected spaces, such that each streamline is fully contained within one simply connected space. Using the continuity equation given by Eq.\ \eqref{eq:stokesb}, there exist scalar functions $\Lambda$ and $X$ such that $\nabla \Lambda \times \nabla X = \vec{u}$ \cite{bear1988dynamics,aris1989vectors}.  The scalar functions $\Lambda$ and $X$ represent two families of stream surfaces whose intersections are the streamlines.

Every point in $x \in \Omega_s$ is uniquely described by the streamline $\mathcal{S}$ passing through $x$, and the distance $s$ along $\mathcal{S}$ from inlet to point $x$. The streamline $\mathcal{S} = \mathcal{S}(\Lambda=\lambda,X=\chi)$ is the intersection of the surfaces given by the constants $\lambda$ and $\chi$. This is similar to a Lagrangian frame of reference, however we use distance instead of time to distinguish points on a streamline. A change of variables from the usual Cartesian coordinates $(x,y,z)$ to the streamline coordinates $(\Lambda,X,s)$ gives the Jacobian
\begin{equation}
  \left\lVert \frac{\delta (\Lambda,X,s)}{\delta (x,y,z)} \right\lVert = (\nabla \Lambda \times \nabla X) \cdot \nabla s = \vec{u} \cdot \frac{\vec{u}}{u} = u.
\end{equation}

By vector calculus identities we have $\vec{u}  = \nabla \Lambda \times \nabla X = \nabla \times \Lambda \nabla X$. Using Stokes' theorem, the fluid discharge $\hat{Q}$ through a streamtube bounded by four stream surfaces represented by the constants $\lambda_1,\lambda_2,\chi_1$ and $\chi_2$ is then
\begin{equation}
\hat{Q} = \int \int_S \vec{u} \cdot \vec{n} dS = \int_{\delta S} \Lambda \nabla X \cdot \vec{l} ds = 
(\lambda_2-\lambda_1)(\chi_2-\chi_1),
\label{eq:streamtube_to_discharge}
\end{equation}
where $S$ is a cross section of the streamtube, and $\vec{l}$ is the unit tangent to the surface boundary $\delta S$. In the last equality we use that either $\Lambda$ or $X$ is constant for each of the four line segments in $\delta S$.

We can now define a permeability factor for the individual streamlines. Starting with Eq.\ \eqref{eq:kappa_glob_def} we have
\begin{align}
\kappa_s &= \frac{1}{\Omega_s} \int_{\Omega_s} \kappa dV = \frac{1}{\Omega_s} \int \int \int_{\mathcal{S}(\Lambda,X)} \frac{\kappa}{u} ds dX d\Lambda \notag \\
 &= \frac{1}{\Omega_s} \int \int \int_{\mathcal{S}(\Lambda,X)} - \mu \rho g \frac{\nabla h \cdot \vec{u}}{u} \left( \frac{\Delta s}{\rho g \Delta h} \right)^2 ds dX d\Lambda \notag \\
 &= \frac{1}{\Omega_s} \int_\mathbb{S} - \mu \rho g \Delta h \left( \frac{\Delta s}{\rho g \Delta h} \right)^2  dQ_\mathcal{S} = \frac{1}{\Omega_s} \int_\mathbb{S} \kappa(\mathcal{S})  dQ_\mathcal{S},
\label{eq:kappa_streamlines} 
\end{align}
where $\kappa(\mathcal{S}) = - \mu \Delta s^2/(\rho g \Delta h)$ is the permeability factor for the streamline $\mathcal{S}$. For the fourth equality we use that $\nabla h \cdot \vec{u} / u = \delta h/\delta s$, and that the total head difference along a streamline is equal the applied head difference $\Delta h$. The infinitesimal fluid discharge for the infinitesimal streamtube given by $dX$ and $d\Lambda$ is denoted by $dQ_\mathcal{S}$, where $dX d\Lambda = dQ_\mathcal{S}$ from Eq.\ \eqref{eq:streamtube_to_discharge}. 

Note that $\kappa(\mathcal{S})$ is a constant, and that
\begin{equation}
\kappa_s =   \frac{1}{\Omega_s} \int_\mathbb{S}  \kappa(\mathcal{S}) dQ_\mathcal{S} = \frac{1}{\Omega_s} \kappa(\mathcal{S}) \int_\mathbb{S}  dQ_\mathcal{S}  = \frac{Q}{\Omega_s} \kappa(\mathcal{S}).
\label{eq:kappa_by_streamlines}
\end{equation}

Rewriting the expression for $\kappa(\mathcal{S})$, we obtain:
\begin{equation}
\kappa(\mathcal{S}) = - \frac{\mu \Delta s^2}{\rho g \Delta h}
= \left( \frac{\Delta s}{l_\mathcal{S}} \right)^2 \int_{\mathcal{S}} -\frac{\mu u}{\rho g \nabla h \cdot \vec{u}} ds \left( \frac{1}{l_\mathcal{S}^2} \Delta h \int_{\mathcal{S}}  \frac{u}{\nabla h \cdot \vec{u}}ds \right)^-,
\label{eq:kappa_streamline_segmented}
\end{equation}
where $l_\mathcal{S}$ is the length of the streamline $\mathcal{S}$. In the following, we will link the permeability factor for a streamline to descriptors of the pore structure: namely, tortuosity, constriction and hydraulic conductance; all represented in Eq.\ \eqref{eq:kappa_streamline_segmented}.

\subsection{Tortuosity}

The \emph{tortuosity} of the streamline $\mathcal{S}$ is given by $\tau(\mathcal{S})=\Delta s/l_\mathcal{S}$, i.e.\ the length of the porous medium divided by the length of the streamline \cite{bear1988dynamics,koponen1996tortuous}.  Due to the linearity of the Stokes equations, Eqs.\ \eqref{eq:stokes}, the streamline $\mathcal{S}$ is independent of the magnitude of applied piezometric head drop $\Delta h$ and the constants $\rho$ and $\mu$, thus $\tau(\mathcal{S})$ is only dependent on pore structure and direction of the applied piezometric head drop.

For smaller tortuosity $\tau(\mathcal{S})$ the fluid needs to travel longer distance, and more applied head potential is expended due to transport distance. This increase in energy expenditure is reflected in the smaller factor $\tau(\mathcal{S})^2=(\Delta s/l_\mathcal{S})^2$ in Eq.\ \eqref{eq:kappa_streamline_segmented}. Longer travel distance for the fluid decreases the effectiveness of the pore space to conduct flow.

\subsection{Constriction Factor}
\label{subsec:streamline_constriction_factor}

For a straight circular pore channel of length $L$ with cross-sectional area $A(x)$ at point $x$, the degree of variation in cross-sectional area can be measured by the constriction factor
\begin{align}
C &= \frac{1}{L^2} \int_0^L A(x)^2 dx \int_0^L \frac{1}{A(x)^2} dx
\label{eq:const_old_def} \\
&= \frac{1}{L^2} \int_0^L \frac{Q}{\rho g \nabla h(x)} dx \int_0^L \frac{\rho g \nabla h(x)}{Q} dx \notag \\
&= \frac{1}{L^2} \int_0^L \frac{1}{\nabla h(x)} dx \int_0^L \nabla h(x) dx, \label{const_new_def}
\end{align}
corresponding to definitions introduced in Refs.\ \cite{dullien1992porous,berg2012}. For the second equality we assume the fluid flow is described by the Hagen-Poiseuille equation (see Eq.\ \eqref{hagen-poiseuille_q}), thus $Q/(\rho g \nabla h(x)) \propto A(x)^2$.  When the fluid is incompressible, the total discharge $Q$ must be constant through all pore channel cross-sections due to mass-balance, yielding Eq.\ \eqref{const_new_def}. For porous media in general, the cross-sectional area $A(x)$ used in Eq.\ \eqref{eq:const_old_def} is not straight-forward defined. As seen in Sect. \ref{subsec:constriction}, $C$ represents the reduction in permeability due to the variation in cross-sectional area.

Following Ref.\ \cite{berg2012} we propose a (hydraulic) \emph{constriction factor} for streamline $\mathcal{S}$ by replacing the head gradient $\nabla h$ in Eq.\ \eqref{const_new_def} with the head derivative $\delta h /\delta s = \nabla h \cdot \vec{u} / u$ along the streamline:
\begin{equation}
C(\mathcal{S}) = \frac{1}{l_\mathcal{S}^2} \int_\mathcal{S} \frac{u}{ \nabla h \cdot \vec{u}} ds \int_\mathcal{S} \frac{ \nabla h \cdot \vec{u} }{ u } ds = \frac{1}{l_\mathcal{S}^2} \Delta h \int_\mathcal{S} \frac{u }{ \nabla h \cdot \vec{u}} ds.
\label{eq:const_one_path}
\end{equation}
As with the tortuosity $\tau(\mathcal{S})$, the constriction factor $C(\mathcal{S})$ is only dependent on pore structure and direction.

When the fluid flows through a constriction, the head derivative $\delta h /\delta s$ along the streamline increases. A large variation in pore size along the streamline then translates into a large variation in the head derivative. The constriction factor $C(\mathcal{S})$ thus relates to the constricting and expanding nature of the pore space along the streamline $\mathcal{S}$, or equivalently the converging-diverging set of streamlines around streamline $\mathcal{S}$. 
For a larger constriction factor $C(\mathcal{S})$ the effectiveness of the pore space to conduct flow is reduced.

\subsection{Hydraulic Conductance}

The microscopic hydraulic conductance is given by 
\begin{equation}
B = -\frac{\mu u^2}{\rho g \nabla h \cdot \vec{u}},
\label{eq:local_hydraulic_conductance}
\end{equation}
and is related to the pore size and shape, and the location in the pore \cite{bear1967generalized,bear1988dynamics}. 
Following Eq.\ \eqref{eq:kappa_streamlines}, the \emph{hydraulic conductance} for a streamline $\mathcal{S}$ is
\begin{equation}
B(\mathcal{S}) =  \int_{\mathcal{S}} B \frac{1}{u} ds =   \int_{\mathcal{S}} -\frac{\mu u}{\rho g \nabla h \cdot \vec{u}} ds.
\end{equation}
Observe that $B(\mathcal{S})$ represents the second term of Eq.\ \eqref{eq:kappa_streamline_segmented}. Also note that $B(\mathcal{S})$ is dependent on both the magnitude of the applied piezometric head drop $\rho g \Delta h$ and the viscosity $\mu$, in addition to pore structure and direction of the applied head drop. This is in contrast with the hydraulic conductance $B$, tortuosity $\tau(\mathcal{S})$ and constriction factor $C(\mathcal{S})$, which are only dependent on pore structure and direction.

With the formulations above, Eq.\ \eqref{eq:kappa_streamline_segmented} can now be rewritten as follows:
\begin{equation}
\kappa(\mathcal{S}) = \frac{B(\mathcal{S}) \tau(\mathcal{S})^2}{C(\mathcal{S})}.
\label{eq:kappa_eq_Btortconst_one_path}
\end{equation}
The permeability factor for an individual streamline is then expressed by descriptors of the pore structure.

\section{Effective Permeability}
\label{sec:perm_description}

In this section we show how the permeability factor $\kappa_s$ can be segmented into a characteristic length, a constriction factor and a tortuosity by averaging over the streamline values. These are strictly related to fluid flow, based on the solution of the Stokes equation inside the porous medium.

The effective hydraulic conductance is found as the volume-weighted average of the hydraulic conductance $B$ \cite{bear1967generalized,bear1988dynamics}:
\begin{equation}
B_s = \frac{1}{\Omega_s} \int_{\Omega_s} B dV.
\label{eq_global_hydraulic_conductance}
\end{equation}
Since $B$ is only dependent on pore space and direction, so is $B_s$. From 
\begin{equation}
B_s = \frac{1}{\Omega_s} \int_{\Omega_s} B dV = \frac{1}{\Omega_s}\int_\mathbb{S} B(\mathcal{S}) dQ_\mathcal{S},
\label{eq_global_hydraulic_conductance}
\end{equation}
 we have a correspondence between the volume integral of $B$ and the streamline integral of $B(\mathcal{S})$.

We define the \emph{characteristic (hydraulic) length} as $L_h = \sqrt{8 B_s}$ to represent the effective hydraulic pore radius of the porous medium. 
Note that the characteristic length scales linearly with the size of the porous medium, as desired for a characteristic length. 
As seen in Sect.\ \ref{subsec:char_length}, for a porous medium consisting of parallel circular tubes of radius $r$ and length $\Delta s$, where these tubes connect the opposite sides of a cube of side length $\Delta s$, then $L_h = r$, and $k = \phi_s \kappa_s = \phi_s B_s = \phi_s L_h^2/ 8$ as desired for such a medium \cite{dullien1992porous}.

Consider a porous medium for which $C(\mathcal{S})=1$ for all streamlines $\mathcal{S}$, e.g.\ a single tube of constant cross-sectional area. Following Eq.\ \eqref{eq:kozeny-carman}, for such a porous medium we desire for a tortuosity $\tau_s^2$ to give $k = \phi_s B_s \tau_s^2$ \cite{dullien1992porous,bear1988dynamics}. Then $\kappa_s = B_s \tau^2_s$ by Eq.\ \eqref{eq:darcy_global}, thus invoking Eq.\ \eqref{eq:kappa_eq_Btortconst_one_path} gives
\begin{equation}
\tau^2_s = \frac{1}{\int_\mathbb{S} B(\mathcal{S}) dQ_\mathcal{S}}\int_\mathbb{S} \tau^2(\mathcal{S}) B(\mathcal{S}) dQ_\mathcal{S}.
\label{eq:tort_path}
\end{equation}
Hence the tortuosity squared $\tau_s^2$ is a weighted average of the streamline tortuosity squared.  Note that the tortuosity $\tau^2_s$ is only dependent on the pore space $\Omega_s$ and the direction of applied piezometric head drop, it is dimensionless and scale invariant.

The tortuosity is commonly formulated as a weighted average of the streamline lengths \cite{duda2011hydraulic,koponen1997permeability}, therefore let the tortuosity $\hat{\tau}^\alpha= 1/\int_\mathbb{S} \hat{w}(\mathcal{S}) dQ_\mathcal{S} \int_\mathbb{S} \tau^\alpha (\mathcal{S}) \hat{w}(\mathcal{S}) dQ_\mathcal{S}$ be another weighted average of streamline tortuosity $\tau(\mathcal{S})$, now to a power $\alpha$. Consider a porous medium for which $C(\mathcal{S})=1$ and $\tau(\mathcal{S})$ is constant for all streamlines $\mathcal{S}$. We still want $k = \phi_s B_s \hat{\tau}^\alpha$, thus $\hat{\tau}^\alpha = \tau_s^2$, which yields 
\begin{equation}
\tau^\alpha(\mathcal{S}) = \frac{1}{\int_\mathbb{S} \hat{w}(\mathcal{S}) dQ_\mathcal{S}} \int_\mathbb{S} \tau^\alpha (\mathcal{S}) \hat{w}(\mathcal{S}) dQ_\mathcal{S} = \hat{\tau}^\alpha = \tau_s^2 = \tau^2(\mathcal{S}),
\end{equation}
therefore $\alpha=2$. 
\sloppy If we desire a streamline decomposition to hold for $B_s \hat{\tau}^2 = \int_\mathbb{S} \kappa(\mathcal{S}) dQ_\mathcal{S}$, then $(\int_\mathbb{S} B(\mathcal{S}) dQ_\mathcal{S} / \int_\mathbb{S} \hat{w}(\mathcal{S}) dQ_\mathcal{S}) \hat{w}(\mathcal{S}) = B(\mathcal{S})$, yielding $\tau_s^2$ unique of the form described by $\hat{\tau}^\alpha$.

Using Eqs.\ \eqref{eq:kappa_by_streamlines}, \eqref{eq:kappa_eq_Btortconst_one_path}, \eqref{eq_global_hydraulic_conductance} and \eqref{eq:tort_path}, we have
\begin{equation}
B_s \tau_s^2 = \frac{1}{\Omega_s} \int_\mathbb{S} B(\mathcal{S}) \tau^2(\mathcal{S}) dQ_\mathcal{S} = \frac{1}{\Omega_s} \int_\mathbb{S} \kappa(\mathcal{S}) C(\mathcal{S}) dQ_\mathcal{S} = \kappa_s \frac{1}{Q} \int_\mathbb{S} C(\mathcal{S}) dQ_\mathcal{S}.
\label{eq:Btort_eq_kappa_const}
\end{equation}
By factoring out the hydraulic conductance $B_s$ and tortuosity $\tau_s^2$, the remaining contribution to the permeability factor $\kappa_s$ is 
\begin{equation}
C_s = \frac{1}{Q} \int_\mathbb{S}   C(\mathcal{S}) dQ_\mathcal{S},
\label{eq:const}
\end{equation}
where $C_s$ is denoted the (hydraulic) \emph{constriction factor.} The constriction factor is also only dependent on the pore space $\Omega_s$ and direction, it is dimensionless and scale invariant.  While the hydraulic conductance represents an effective hydraulic pore radius, the constriction factor represents the fluctuation in hydraulic pore radii.

From Eq.\ \eqref{eq:Btort_eq_kappa_const} and Eq.\ \eqref{eq:const} we have
\begin{equation}
\kappa_s = \frac{B_s \tau^2_s}{C_s} = \frac{\tau_s^2 L_h^2}{8 C_s}.
\label{eq:kappa_eq_Btortconst_global}
\end{equation}
Combining Eq.\ \eqref{eq:darcy_global} with Eq.\ \eqref{eq:kappa_eq_Btortconst_global} then gives:
\begin{equation}
k = \kappa_s \phi_s = \frac{\tau_s^2 B_s \phi_s}{C_s} = \frac{\tau_s^2 L_h^2 \phi_s}{8 C_s}.
\label{eq:perm_as_Btort_const}
\end{equation}
We thus have a full description of the porous medium permeability by pore structure related parameters.

The permeability factor $\kappa_s$ gives the effectiveness of the pore space $\Omega_s$ to conduct flow. This effectiveness is reduced by longer flow paths given with a smaller tortuosity $\tau_s$, more variation in pore size along the flow paths described by a larger constriction factor $C_s$, and smaller pores reflected by a smaller characteristic length $L_h$. Note that these factors are dependent on direction in addition to the pore structure, which leads to the anisotropy of the permeability.

\section{Single Tube Example}
\label{sec:hagen-poiseuille}

Capillary bundle of tube models have a wide use as simplified representations of porous media. The single tube examples in this section illustrate such model representations.

Consider a straight cylindrical tube with constant cross-section of radius $R$. If the length of the tube is much larger than the radius, then the flow inside the tube is approximated by the Hagen-Poiseuille equation 
\begin{equation}
Q= -\frac{\pi R^4 \rho g \nabla h}{8 \mu}.
\label{hagen-poiseuille_q}
\end{equation}
Moreover, the flow velocity is given by 
\begin{equation}
u(r) =-\frac{(R^2-r^2) \rho g \nabla h}{4 \mu},
\label{hagen_poiseuille_u}
\end{equation}
where $r$ is the distance from the center of the tube \cite{adler1992porous}. In the equation above and subsequently in this section, $\nabla h$ is also used to denote the scalar value $-\lVert \nabla h \rVert$. 
In the following subsections, the Hagen-Poiseuille equation is used to demonstrate our theory introduced above.

\subsection{Permeability Factor}

Consider a straight cylindrical tube inside a cube of side-length $L$, and with an applied piezometric head difference $\Delta h$ over two opposite sides of the cube. Let the tube be of length $L$ and aligned with the applied head, then the piezometric head gradient inside the tube is $\nabla h = \Delta h/L$.

Combining Darcy's law with the Hagen-Poiseuille equation, Eqs.\ \eqref{eq:darcy} and \eqref{hagen-poiseuille_q}, the permeability of the cube is given by:
\begin{equation}
k = -\frac{Q \mu L}{L^2 \rho g \Delta h} = \frac{\pi R^4}{8 L^2}.
\label{hagen-poiseuille_darcy}
\end{equation}

Using the flow velocity a distance $r$ from the tube center as given by Eq.\ \eqref{hagen_poiseuille_u}, and the piezometric head gradient $\nabla h = \Delta h/L$, the permeability factor described in Eq.\ \eqref{eq:kappa_def} is $\kappa(r) = (R^2-r^2)/4$. Taking the volume-weighted average then gives the effective permeability factor
\begin{equation}
\kappa_s = \frac{1}{\pi R^2 L} \int_0^L \int_0^R \kappa(r) 2\pi r dr ds = \frac{R^2}{8}.
\label{eq:int_of_kappa}
\end{equation}
This gives $\kappa_s \phi =  \pi R^4/(8 L^2)$, which is equal to the result from Darcy's law in Eq.\ \eqref{hagen-poiseuille_darcy}, hence our results are consistent with Eq.\ \eqref{eq:darcy_global}.

\subsection{Characteristic Length}
\label{subsec:char_length}

Let our porous medium and applied piezometric head be as above. The permeability can be calculated using the individual contributions from the characteristic length, constriction, and tortuosity, as given by Eq.\ \eqref{eq:perm_as_Btort_const}. The tortuosity and constriction are equal to $1$ when the fluid flow inside the tube is described by the Hagen-Poiseuille equation, while the hydraulic conductance is $B(r) = (R^2-r^2)/4$. The volume-weighted average of $B(r)$ is then equal to the integration of $\kappa(r)$ in Eq.\ \eqref{eq:int_of_kappa}, therefore $B_s = R^2/8$. This gives a characteristic length $L_h = \sqrt{8B_s} = R$. The characteristic length thus equals the radius of the tube, as desired. Since $\tau_s = 1$ and $C_s = 1$, we have
\begin{equation}
\frac{\tau^2_s L_h^2 \phi}{8C_s} = \frac{\pi R^4}{8 L^2},
\end{equation}
which agrees with the result from Darcy's law in Eq.\ \eqref{hagen-poiseuille_darcy}, hence our results are consistent with Eq.\ \eqref{eq:perm_as_Btort_const}.

\subsection{Tortuosity}

\begin{figure}
\centering
\begin{minipage}[b]{0.45\linewidth}
\includegraphics[width=\linewidth]{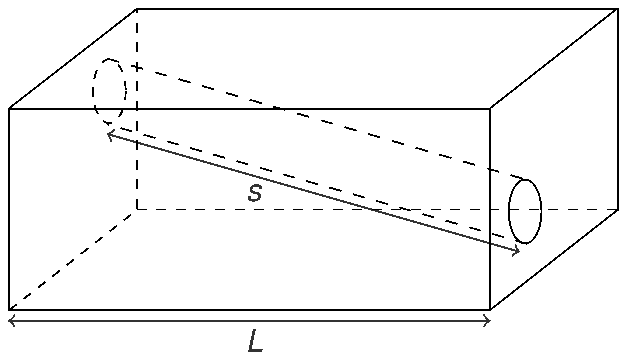}
\caption{Idealized porous medium with tortuosity different from $1$.}
\label{tort_tube_img}
\end{minipage}
\quad
\begin{minipage}[b]{0.45\linewidth}
\includegraphics[width=\linewidth]{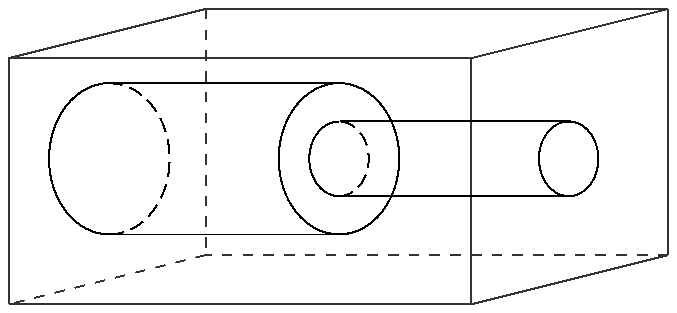}
\caption{Idealized porous medium with a single constriction.}
\label{constriction_img}
\end{minipage}
\end{figure}

Let again our porous medium and applied piezometric head be as above, except that the length of the straight circular tube with constant cross-sectional area is $s > L$, as shown in Fig.\ \ref{tort_tube_img}. 
Then the piezometric head gradient inside the tube is $\nabla h = \Delta h/s$.

Combining Eqs.\ \eqref{eq:darcy} and \eqref{hagen-poiseuille_q} gives the permeability as:
\begin{equation}
k = -\frac{Q \mu L}{L^2 \rho g \Delta h} = \frac{\pi R^4}{8 Ls}.
\label{hagen-poiseuille_darcy_tort}
\end{equation}
Note that a larger $s$ gives a smaller permeability $k$, while $s=L$ gives a permeability equal to Eq.\ \eqref{hagen-poiseuille_darcy}.

The calculations for the characteristic length in Sect.\ \ref{subsec:char_length} still hold, so $L_h = R$. The constriction factor $C_s$ is equal to $1$, while each streamline has length $s$, thus $\tau_s=\tau(\mathcal{S})=L/s$. 
We then have 
\begin{equation}
\frac{\tau^2_s L_h^2 \phi}{8C_s} = \frac{\pi R^4}{8 Ls},
\end{equation}
which is equal to the result from Darcy's law in Eq.\ \eqref{hagen-poiseuille_darcy_tort}, and consistent with Eq.\ \eqref{eq:perm_as_Btort_const}.

Note that by changing the orientation of the tube shown in Fig.\ \ref{tort_tube_img} the tortuosity and the permeability will change, while in this idealized case the characteristic length and constriction factor stay constant. In this case the anisotropy of the permeability is captured by the tortuosity.

\subsection{Constriction}
\label{subsec:constriction}

We will now use an idealized porous medium to investigate the effect of constriction. Consider a porous medium consisting of two tube segments in sequence aligned with the applied piezometric head, as depicted in Fig.\ \ref{constriction_img}. The two tube segments both have length $L/2$ inside a cube of side-length $L$, while the radii of the two segments are $R_1$ and $R_2$. 
When assuming $R_i \ll L/2$ for $i=1,2$, the flow inside the tube segments can be approximated by the Hagen-Poiseuille equations and the tortuosity can be approximated as $\tau_s = 1$.

Invoking Eq.\ \eqref{hagen-poiseuille_q} we can show that the discharge is $Q = -(\pi R_1^4 R_2^4 \rho g \nabla h)/(4 \mu  (R_1^4 + R_2^4)).$ 
Using Darcy's law, Eq.\ \eqref{eq:darcy}, we then obtain
\begin{equation}
k= - \frac{Q\mu L}{L^2 \rho g \Delta h} = \frac{\pi R_1^4 R_2^4}{4L^2 (R_1^4 + R_2^4)}.
\label{hagen-poiseuille_darcy_const}
\end{equation}

From Eq.\ \eqref{hagen-poiseuille_q} we have $\nabla h_i = -8 \mu Q /(\pi R_i^4 \rho g)$, and from Eq.\ \eqref{eq:const_one_path} 
we derive the constriction factor as
\begin{equation}
C_s = \frac{1}{L^2} \int_0^L \nabla h dx \int_0^L \frac{dx}{\nabla h} = \frac{1}{4}\frac{ (R_1^4 + R_2^4)^2}{R_1^4 R_2^4}.
\end{equation}
For each tube section we have $B_i = R_i^2/8$, which gives $B_s= (R_1^4+R_2^4)/(8(R_1^2 + R_2^2))$. Since $\tau_s^2 = 1$, we have:
\begin{equation}
k = \frac{\tau_s^2 B_s \phi_s}{C_s} = \frac{\pi R_1^4 R_2^4}{4L^2 (R_1^4 + R_2^4)},
\end{equation}
consistent with Eq.\ \eqref{hagen-poiseuille_darcy_const}. Note that a larger difference between $R_1$ and $R_2$ gives a larger constriction factor $C_s$, which then implies a lower permeability, while $C_s=1$ when $R_1=R_2$, as desired.

We will now revisit the more general constriction example from Sect.\ \ref{subsec:streamline_constriction_factor}, where we considered a tube with cross-sectional area $A(x)$ for $x \in [0,L]$. We still assume that the flow inside the tube is approximated by the Hagen-Poiseuille equations and that the tortuosity can be approximated as $\tau_s = 1$. Then $\nabla h (x) = -8 \mu Q \pi /(A(x)^2 \rho g)$, from Eq.\ \eqref{eq:darcy} we then have
\begin{equation}
k = \frac{1}{8 \pi L \int_0^L \frac{1}{A(x)^2}dx}.
\end{equation}

The constriction factor
\begin{equation*}
C_s = \frac{1}{L^2} \int_0^L \frac{1}{A(x)^2} dx \int_0^L A(x)^2 dx
\end{equation*}
equals Eq.\ \eqref{eq:const_old_def} in Sect.\ \ref{subsec:streamline_constriction_factor}. For the cross-section at point $x$ we have $B(x) = A(x)/(8 \pi)$, thus $B_s = \int A(x)^2 dx/(8 \pi \int A(x) dx)$.

Now assume another porous medium with equal pore volume and with a constant cross-sectional area. Then the cross-sectional area is $\hat{A}=(1/L) \int A(x) dx$, $\hat{B}_s = \hat{A}/(8\pi)$ and $\hat{k} = \hat{A}^2/(L^2 8 \pi)$. When factoring out the hydraulic conductance, the reduction in permeability due to the varying cross-sectional area is
\begin{equation}
\frac{k/B_s}{\hat{k}/\hat{B}_s} =  \frac{1}{C_s},
\end{equation}
hence the reduction equals the inverse of the constriction factor.

\section{Fontainebleau Rock Example}\label{section:bentheimer_network}

We next turn to natural porous media, such as given by micro-CT (microtomography) images and rock models of Fontainebleau sandstone. Using the e-Core software \cite{e-core_v152} we generated three-dimensional rock models of Fontainebleau sandstone with porosities ranging from 8 to 26\%. We used the exact same grain packing for all models, while we changed the amount of quartz cementation to achieve the variation in porosities. The rock modeling process is described in detail in Refs.\ \cite{oren2006digital,berg2012}. The micro-CT images and model sample size were 2.7 mm cubed with a resolution of 5.7 $\mu$m.

\begin{figure}
\centering
\includegraphics[width=7cm]{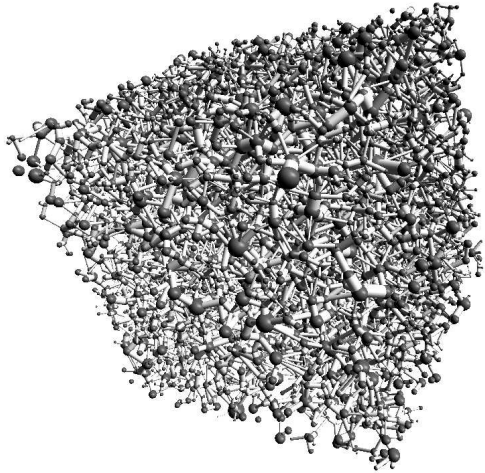}
\caption{Visualization of the network representation of the pore space of Fontainebleau sandstone mCTc in Tabel \ref{tab_models}. Balls represent pore bodies, and sticks represent pore throats. This network has 7413 pore bodies and 14254 pore throats.}
\label{network_img}
\end{figure}

We extracted network analogs using the e-Core software \cite{e-core_v152}. The software extracts a pore network using a grain based algorithm \cite{bakke19973}, which segments the pore space into pore bodies and pore throats, each associated with a point $x \in V$, a volume, a shape factor $G$ \cite{mason1991capillary}, and an inscribed radius $r$ \cite{patzek2001shape}. One such network is visualized in Fig.\ \ref{network_img}. The distance between the two pore bodies $t1$ and $t2$ connected by a pore throat $t3$ is divided into parts: parts $l_{t1}$ and $l_{t2}$ are associated with each pore body, $t1$ and $t2$, respectively; and part $l_{t3}$ associated with the pore throat.

Following Eq.\ (18) in Ref.\ \cite{oren1998extending}, the hydraulic conductance of each part is approximated by
\begin{equation}
g = \frac{3 r^4}{80 \mu G}.
\label{eq:hydraulic_conductance}
\end{equation}
The effective hydraulic conductance between two pore bodies $t1$ and $t2$ connected by a pore throat $t3$ is taken as the length-weighted harmonic average of the three parts
\begin{equation}
g_t = l_t \left( \frac{l_{t1}}{g_{t1}} + \frac{l_{t3}}{g_{t3}} + \frac{l_{t2}}{g_{t2}} \right)^{-1},
\end{equation}
where $l_t = l_{t1}+l_{t3}+l_{t2}$. Assuming a Hagen-Poiseuille type relation between the fluid discharge $Q_{t_{1,2}}$ from pore $t1$ to pore $t2$ and the head gradient $\nabla h_{t_{1,2}} = (h_{t2} - h_{t1})/l_t$ for each pore throat $t$, we have
\begin{equation}
Q_{t_{1,2}} = -g_t \frac{\rho g(h_{t2}-h_{t1})}{l_t},
\end{equation}
where $h_{ti}$ is the piezometric head associated with the pore body $ti$.

\begin{table*}
\begin{center}
\caption{Micro-CT and model results.}
\label{tab_models}
\begin{tabular}{|c|c|c|c|c|c|c|c|c|c|c|}
\hline
&\multicolumn{2}{c|}{Porosity}&$\kappa_s$&\multicolumn{2}{c|}{Tortuosity}&Const.&\multicolumn{2}{c|}{Char.length}&\multicolumn{2}{c|}{Permeability}  \\
Rock&$\phi$&$\phi_s$&$[(\mu m)^2]$&$\tau_s$&$\tau$&$C_s$&$L_h [\mu m]$&$r_c [\mu m]$&$[(\mu m)^2]$&$[mD]$\\
\hline
mCTa	&	0.081		&	0.048	&	0.86	&	0.347	&	0.373	&	107.97	&	78.62	&	10.4	&	0.041	&	41.72	\\
mCTb	&	0.128		&	0.114	&	4.569	&	0.41	&	0.425	&	28.41	&	78.5	&	15.1	&	0.521	&	528.06	\\
mCTc	&	0.176		&	0.166	&	13.165	&	0.449	&	0.462	&	19.91	&	102.06	&	21.3	&	2.192	&	2220.86	\\
mCTd	&	0.21		&	0.2	&	14.078	&	0.447	&	0.464	&	20.49	&	107.39	&	20.3	&	2.811	&	2848.57	\\
a	&	0.086		&	0.056	&	0.766	&	0.337	&	0.355	&	105.3	&	75.28	&	9.1	&	0.043	&	43.44	\\
b	&	0.101		&	0.079	&	1.865	&	0.363	&	0.381	&	58.97	&	81.73	&	13.2	&	0.147	&	149.22	\\
c	&	0.125		&	0.111	&	3.841	&	0.387	&	0.405	&	40.51	&	91.08	&	15.3	&	0.427	&	432.61	\\
d	&	0.153		&	0.143	&	9.573	&	0.431	&	0.451	&	26	&	103.6	&	18.8	&	1.371	&	1388.87	\\
e	&	0.176		&	0.168	&	13.987	&	0.442	&	0.462	&	22.75	&	114.03	&	22.9	&	2.345	&	2376.21	\\
f	&	0.206		&	0.198	&	22.428	&	0.461	&	0.48	&	18.36	&	124.6	&	26.3	&	4.449	&	4507.45	\\
g	&	0.245		&	0.237	&	33.435	&	0.467	&	0.487	&	15.81	&	139.26	&	30.8	&	7.935	&	8040.4	\\
\hline
\end{tabular}
\end{center}
\end{table*}

The network model can now be viewed as a resistor network analog, with a one-to-one correspondence between the pore throats in the porous medium and the resistors in the resistor network analog, and also there is a one-to-one correspondence between the pore bodies and the network nodes. Each pore throat (resistor) $t$ is given a conductance $g_t/l_t$.
Let $h_i$ be the piezometric head corresponding to pore body (node) $i$, and $\{ t_{ij} \}_{j=1}^{\alpha_i}$ the pore throats (resistors) connected to pore body $i$. We then solve for $h_i$ such that 
\begin{equation}
\sum_{j=1}^{\alpha_i}{ -g_t\frac{\rho g(h_j-h_i)}{l_t} } = \sum_{j=1}^{\alpha_i}{ Q_{t_{ij}}} = 0,
\end{equation}
where we have fixed piezometric head at the inlet and outlet boundaries.

\begin{figure}
\centering
\includegraphics[width=7cm]{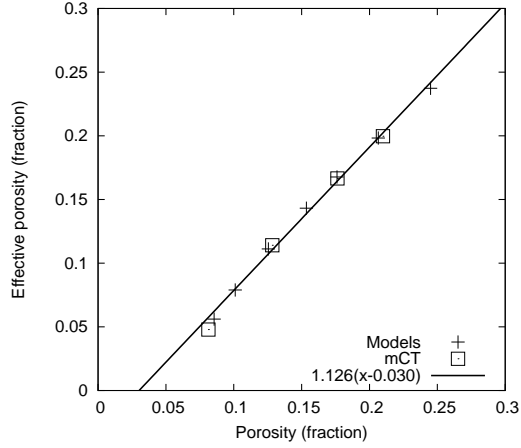}
\caption{Plot showing the correspondence between porosity $\phi$ and $\phi_s$ for the rock models and the micro-CT (mCT) data, together with a linear fit to the model data.}
\label{por_vs_gradpor}
\end{figure}

The network volume with non-zero head gradient can be calculated as
\begin{equation}
\Omega_s = \sum_{h_{t1} \not= h_{t2}}{ \left( \frac{V_{t1}}{\alpha_{t1}} + \frac{V_{t2}}{\alpha_{t2}} + V_{t3} \right) }.
\end{equation}
The values for porosity $\phi$ and $\phi_s = \Omega_s/V$ for the network representations of our micro-CT images and models are reported in Table \ref{tab_models}. In Fig.\ \ref{por_vs_gradpor} we have plotted porosity $\phi$ versus $\phi_s$. A linear fit to the model data plotted in Fig.\ \ref{por_vs_gradpor} gives the correspondence 
\begin{equation}
\phi_s(\phi) = 1.126(\phi - 0.030).
\label{phi_path_to_phi}
\end{equation}

The fluid velocity inside the network elements is not resolved; we therefore threat the fluid velocity as constant inside each network element. The average fluid velocity for part $t1$ and $t2$ are given by $u_{ti} = g_{ti} \rho g \lvert h_{ti}-h_{it} \rvert \alpha_{ti} / V_{ti}$ where $i=1,2$, while for part $t3$ it is $u_{t3} = g_{t3} \rho g \lvert h_{1t}-h_{2t} \rvert / V_{t3}$. Here $h_{1t}$ and $h_{2t}$ are piezometric heads such that $g_{t1} \rho g \lvert h_{t1}-h_{1t} \rvert/l_{t1} = g_{t3} \rho g \lvert h_{1t}-h_{2t} \rvert /l_{t3} = g_{t2} \rho g \lvert h_{2t} -h_{t2} \rvert/l_{t2}$. The fluid velocity $\vec{u}$ is in the opposite direction of the gradient of the head, i.e.\ $\nabla h \cdot \vec{u} = - \lVert \nabla h \rVert u$. The local permeability factors for the sections $t_1, t_2, t_3$, as given by Eq.\ \eqref{eq:kappa_def}, are:
\begin{align}
\kappa_{ti} &= \frac{g_{ti} (\rho g (h_{ti}-h_{it}))^2 \alpha_{ti}}{l_{ti} V_{ti}} \left(\frac{ \Delta s}{\Delta h}\right)^2 \text{ for } i=1,2, \text{ and} \notag \\
\kappa_{t3} &= \frac{g_{t3} (\rho g (h_{1t}-h_{2t}))^2}{l_{t3} V_{t3}} \left(\frac{ \Delta s}{\Delta h}\right)^2. \notag
\end{align}
This enables the calculation of the effective permeability factor as the volume average of these local contributions:
\begin{equation}
\kappa_s = \frac{1}{\Omega_s} \sum_{h_{t1} \not= h_{t2}}{ \left( \frac{V_{t1}}{\alpha_{t1}} \kappa_{t1} + \frac{V_{t2}}{\alpha_{t2}} \kappa_{t2} + V_{t3} \kappa_{t3} \right) }.
\end{equation}
The results are reported in Table \ref{tab_models}.

\begin{figure}
\centering
\begin{minipage}[b]{0.45\linewidth}
\includegraphics[width=0.9\linewidth]{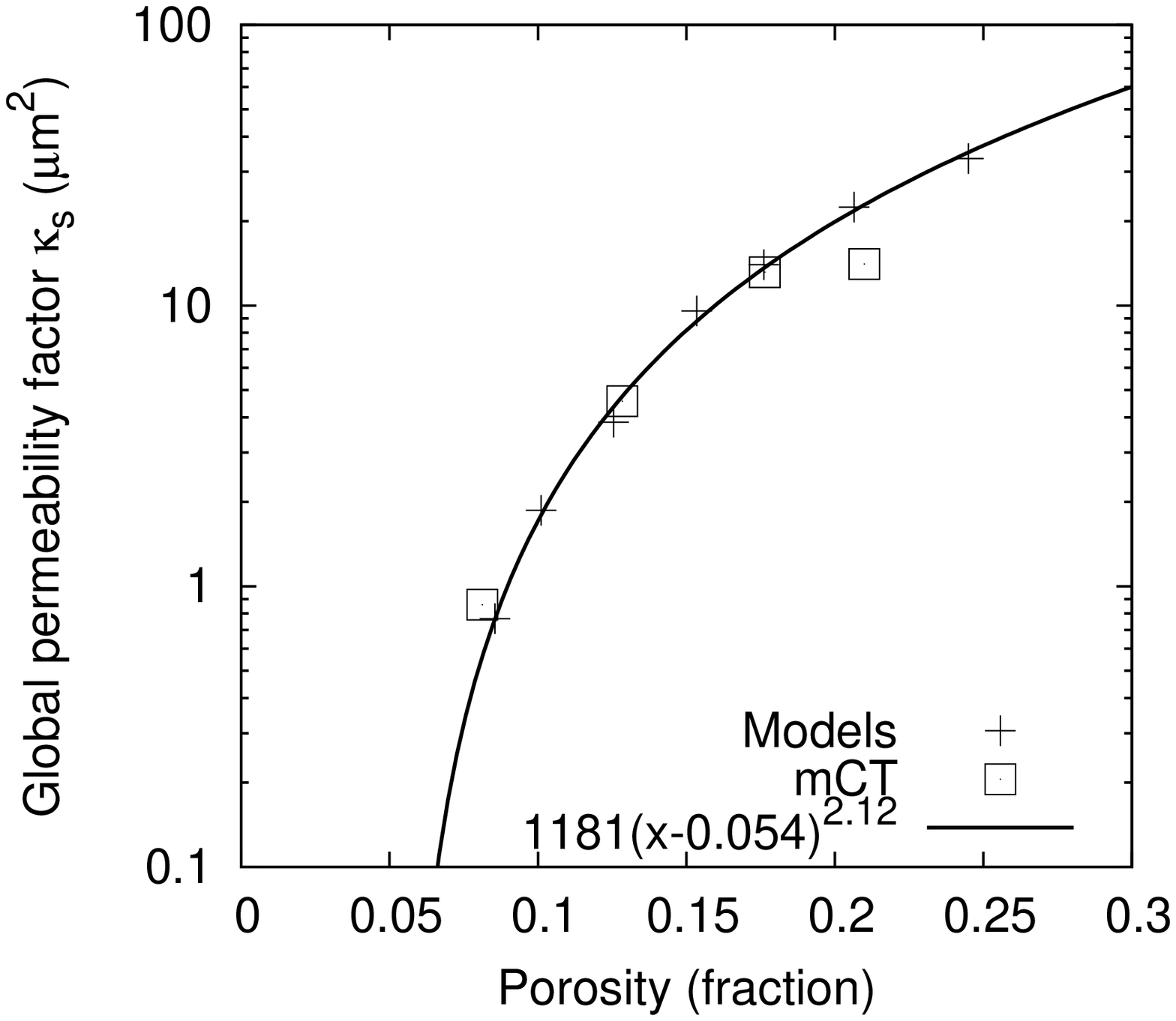}
\caption{Plot showing effective permeability factor $\kappa_s$ versus porosity $\phi$ for the Fontainebleau rock models and the micro-CT (mCT) data, together with a fit to the data.}
\label{por_vs_kappa}
\end{minipage}
\quad
\begin{minipage}[b]{0.45\linewidth}
\includegraphics[width=0.9\linewidth]{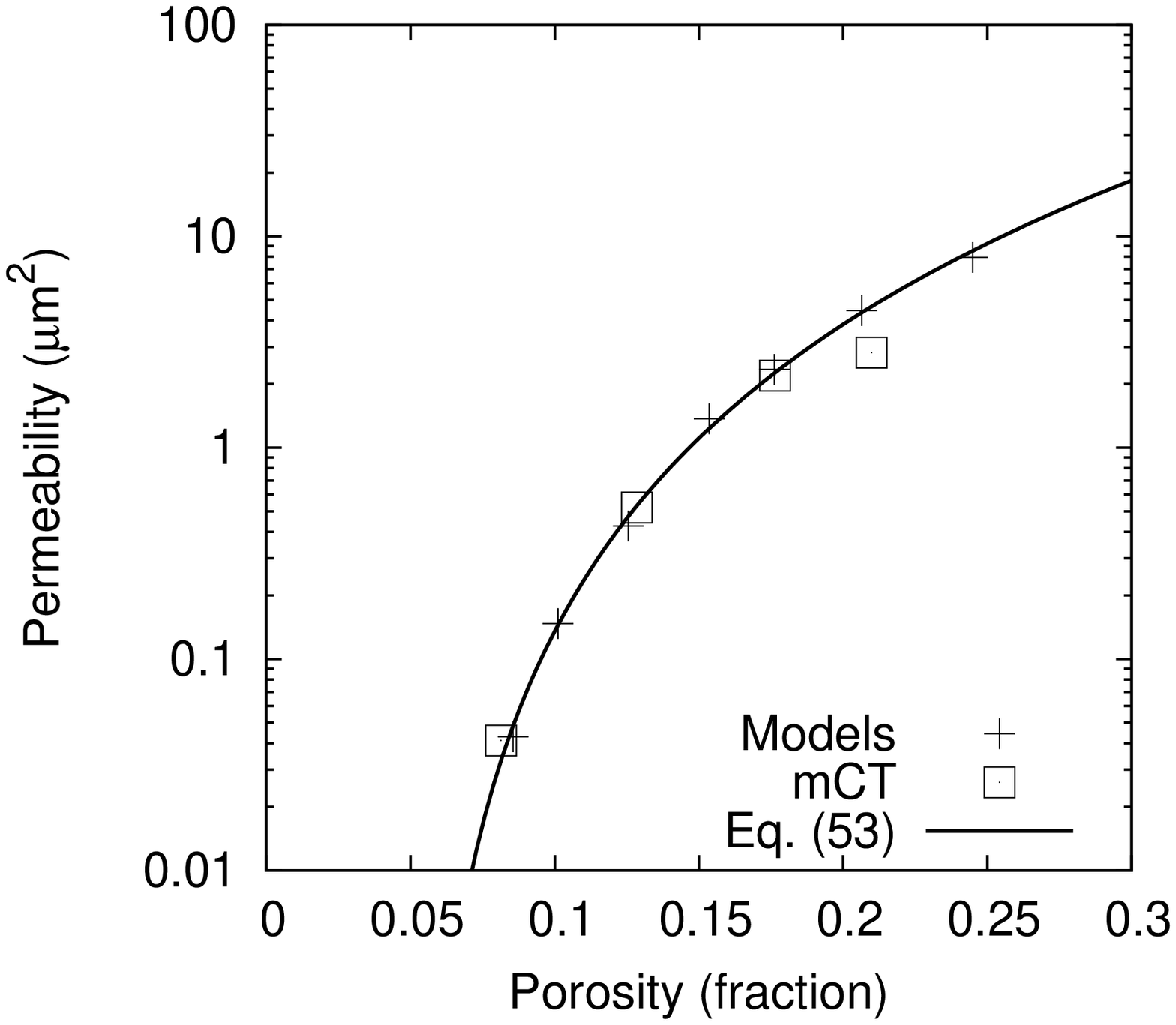}
\caption{Plot showing porosity $\phi$ versus permeability $k$ for the Fontainebleau rock models and micro-CT (mCT) data, together with Eq.\ \eqref{eq_darcy_is_por_and_kappa} describing the correlation.}
\label{por_vs_perm}
\end{minipage}
\end{figure}

The effective permeability factor $\kappa_s$ versus porosity $\phi$ is plotted in Fig.\ \ref{por_vs_kappa}. The function 
\begin{equation}
\kappa_s(\phi)=1181(\phi-0.054)^{2.12},
\label{eq_por_vs_kappa}
\end{equation}
is included as a fit to the model data. For $\phi = 0.054$ we have $\kappa_s(\phi) = 0$, which is interpreted as percolation threshold for this sandstone \cite{mavko1997effect}.

The permeability $k = \kappa_s \phi_s$, as given by Eq.\ \eqref{eq:darcy_global}, is also listed in Table \ref{tab_models}. Calculated porosity and permeability for the network representations of Fontainebleau sandstone are plotted in Fig.\ \ref{por_vs_perm}. Combining Eqs. \eqref{eq:darcy_global}, \eqref{phi_path_to_phi} and \eqref{eq_por_vs_kappa}, we have:
\begin{align}
k(\phi) &=  \kappa_s(\phi) \phi_s(\phi) = 1181(\phi-0.054)^{2.12} 1.126(\phi - 0.030)   \notag \\ 
&=1329 (\phi-0.054)^{2.12} (\phi-0.030).
\label{eq_darcy_is_por_and_kappa}
\end{align}
This function is also included in Fig.\ \ref{por_vs_perm}, and provides a derived porosity-permeability relationship for the Fontainebleau samples.

We discretized the volume $\Omega_s$ into a disjoint union $\sqcup{\mathcal{S}}$, where $\mathcal{S}$ is a subvolume of a single series of pore throats $t_\mathcal{S} = \{ t \}$, with the first throat connected to an inlet boundary and the last connected to an outlet boundary. Each $\mathcal{S}$ transports a constant discharge $Q_\mathcal{S}$, and $\sum{Q_\mathcal{S}} = Q$.  The discretization $\sqcup \mathcal{S}$ is a simplification of the streamlines, similar to the concept of a bundle of capillary tube model. In the network representation for the Fontainebleau sandstone the discretization $\Omega_s = \sqcup{\mathcal{S}}$ is dependent on the fluid flow across the network nodes, however different ways of tracing streamlines across the network nodes were tested, yielding comparable results.

For the sections $t_1, t_2$ and $t_3$ of constant hydraulic conductance as associated with pore throat $t \in t_\mathcal{S}$, we have the corresponding parts $\{ \mathcal{S}_{ti} \}$ of $\mathcal{S}$. Each volume $\mathcal{S}_{ti} \subset V_{ti}$ then has the associated length $l_{ti}$.

The three sections have
\begin{align}
B_{ti} &= -\frac{u}{\rho g \nabla h} = \frac{\mu g_{ti} l_{ti} \alpha_{ti}}{ V_{ti}} \text{ for } i=1,2, \text{ and} \notag \\
B_{t3} &= -\frac{u}{\rho g \nabla h} = \frac{\mu g_{t3} l_{t3}}{ V_{t3}}. \notag
\end{align}
For each $\mathcal{S}$ in $\Omega_s = \sqcup{\mathcal{S}}$ we calculated \begin{equation}
B(\mathcal{S})= \frac{1}{\mathcal{S}} \sum_{t \in t_\mathcal{S}}{\mathcal{S}_{t1} B_{t1} + \mathcal{S}_{t2} B_{t2} + \mathcal{S}_{t3} B_{t3}}.
\end{equation}

We separately calculated the constriction factor
\begin{equation}
C(\mathcal{S}) = \frac{\Delta h }{\left(\sum_{t \in t_\mathcal{S}}{l_t } \right)^2}
 \times \sum_{t \in t_\mathcal{S}}{ \left( \frac{l_{t1}^2}{ \lvert h_{t1}-h_{1t} \rvert }+ \frac{l_{t2}^2}{ \lvert h_{t2}-h_{2t} \rvert} + \frac{l_{t3}^2}{ \lvert h_{1t}-h_{2t} \rvert  }\right) },
\label{eq:const_net_path}
\end{equation}
and the tortuosity
\begin{equation}
\tau(\mathcal{S}) = \frac{\Delta s}{\sum_{t \in t_\mathcal{S}}{ l_t}}.
\end{equation}

We then obtain the characteristic length, constriction factor and tortuosity for the volume $\Omega_s$ as:
\begin{align}
L_h &= \sqrt{8 B_s} = \sqrt{8 \frac{1}{\Omega_s} \sum{\mathcal{S} B(\mathcal{S})} },
\label{eq:B_net_total} \\
C_s & = \frac{1}{Q} \sum{Q_\mathcal{S} C(\mathcal{S})},
\label{eq:const_net_total} \\
\tau^2_s & = \frac{1}{\sum \mathcal{S}B(\mathcal{S})} \sum{ \tau^2(\mathcal{S}) \mathcal{S}B(\mathcal{S})}.
\label{eq:tort_net_total} 
\end{align}
The calculated values are reported in Table \ref{tab_models}. We see that $\tau^2_s L_h^2 / (8 C_s) = \kappa_s$, which is consistent with Eq.\ \eqref{eq:perm_as_Btort_const}.

\begin{figure}
\centering
\includegraphics[width=10cm]{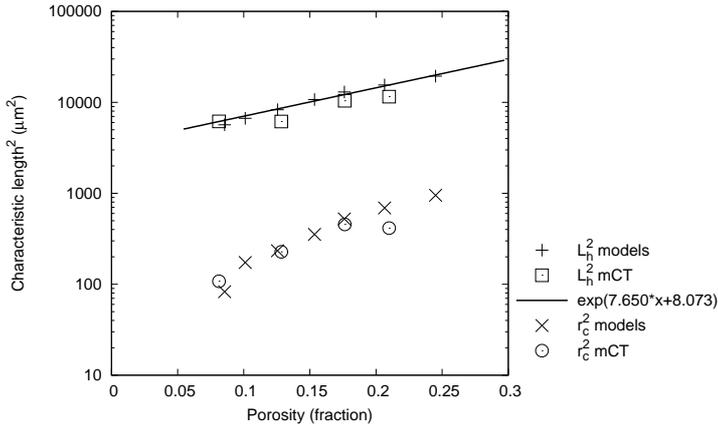}
\caption{Plot showing porosity $\phi$ versus characteristic length squared $L_h^2$ and critical pore radius squared $r_c^2$ for the Fontainebleau rock models and the micro-CT (mCT) data.}
\label{por_vs_char_length}
\end{figure}

For comparison we calculated the tortuosity $\tau$ as
\begin{equation}
\tau =\frac{\Delta s}{ \frac{1}{Q} \sum_t Q_t l_t },
\end{equation}
where $Q_t$ is the discharge through pore throat $t$ \cite{bear1988dynamics,duda2011hydraulic}. Note that the values for $\tau_s$ and $\tau$ are similar, however $\tau_s < \tau$. 
We also calculated the critical pore radius $r_c$ corresponding to the smallest network element radius of the set of largest network elements that percolate through the network \cite{katz1986quantitative}, where the radius of a network element is given by $r\sqrt[4]{3/(10 G \pi)}$. The values are reported in Table \ref{tab_models}. Such characteristic length scales are seen to be significantly lower than the hydraulic characteristic lengths $L_h$ calculated according to Ref.\ \cite{bear1967generalized,bear1988dynamics} for the Fontainebleau networks, included in Fig.\ \ref{por_vs_char_length}.

In Fig.\ \ref{por_vs_char_length} we have plotted porosity $\phi$ versus the characteristic length squared $L_h^2$, the critical pore radius $r_c$, together with a functional fit $L_h^2(\phi) = \exp(7.650\phi+8.073)$. For high porosities, i.e.\ rocks with little cementation and then larger pores, we have larger characteristic length $L_h$ than for low porosities. The characteristic lengths $L_h$ for the micro-CT images scatter around the trend given by the simulated rock models.

\begin{figure}
\centering
\includegraphics[width=10cm]{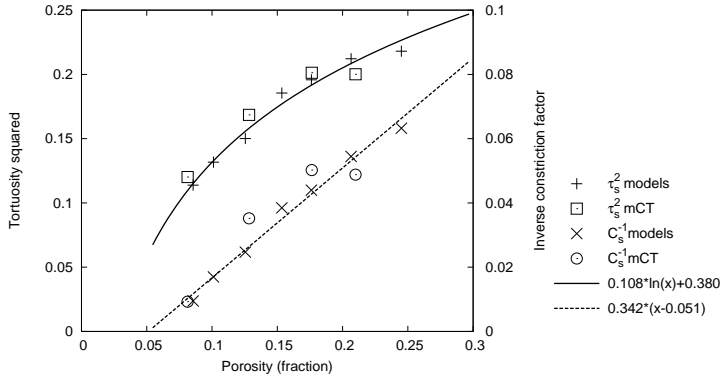}
\caption{Plot showing porosity $\phi$ versus both tortuosity squared $\tau_s^2$ and inverse constriction $C_s^-$ for the Fontainebleau rock models and the micro-CT (mCT) data.}
\label{por_vs_tort_const}
\end{figure}

In Fig.\ \ref{por_vs_tort_const} we have plotted porosity $\phi$ versus both the tortuosity $\tau_s^2$ and the inverse constriction factor $C_s^-$. For high porosities we have less permeability reduction due to both tortuosity and constriction, compared to low porosities. Cementation increases the ratio between pore body and pore throat cross-sectional area, which yields a larger fluctuation in pore size along the streamlines and therefore a higher constriction factor. When cementation blocks pore throats completely, it increases the length of the streamlines and reduces the tortuosity $\tau_s$.

A function $\tau^2_s(\phi) = 0.108\ln(\phi)+0.380$ gives a visual match to the tortuosity of the Fontainebleau sandstone models, while the constriction factor of the models follows a trend $C^-_s(\phi) = 0.342(\phi-0.051)$. Both functions are also plotted in Fig.\ \ref{por_vs_tort_const}. The calculated tortuosity $\tau_s^2$ and inverse constriction $C_s^-$ of the micro-CT images follow the trend given by the models.

The calculated pore structure descriptors for each Fontainebleau sandstone sample reveal a strong functional relation with respect to porosity. This is desirable for sensitive descriptors. The functional trends display a non-trivial behavior at the percolation threshold derived in Eq.\ \eqref{eq_por_vs_kappa}: The tortuosity $\tau_s^2$ and characteristic length $L_h^2$ indicate a non-zero value at the percolation threshold, while the inverse constriction factor $C_s^-$ tends to zero at the percolation threshold.

\section{Conclusion}

In this work we have fundamentally described and calculated the permeability in porous media. The permeability $k$ of a porous medium is equal to $\kappa_s \phi_s$. Here the effective porosity $\phi_s$ is the fractional volume conducting flow from inlet to outlet. An effective permeability factor $\kappa_s$ is given by the volume-weighted average of the microscopic permeability factors
$$\kappa =  -\mu \rho g \nabla h \cdot \vec{u} \left( \frac{\Delta s}{\rho g \Delta h} \right)^2.$$
This microscopic permeability factor $\kappa$ relates the local contribution of the pore structure to effectiveness of the pore space to conduct fluid flow $\kappa_s$.

We have shown that $\kappa_s = \tau_s^2 B_s/C_s = \tau_s^2 L_h^2/(8C_s)$, where the effective pore radius in the porous medium is described by the characteristic length $L_h$, fluctuation in local hydraulic radii is described by the constriction factor $C_s$, and the effective length of the streamlines is described by the tortuosity $\tau_s$. These characteristic length, constriction factor and tortuosity are direction dependent intrinsic descriptors of the pore structure. Their directional dependence leads to anisotropy of the permeability, i.e., the tensorial form of the permeability.

We have shown that our methodology reproduces results for Hagen-Poiseuille flow in tubes. It is also applied to a natural porous medium given by a pore network representation of Fontainebleau sandstone, where we show how the distinct contributions to the permeability from characteristic length, constriction and tortuosity correlate with porosity. As long as the flow and piezometric head field can be obtained, this methodology is applicable to any porous medium.

This work demonstrates how the permeability can be related to porosity, in the sense of Kozeny-Carman, through fundamental and measurable descriptors of the pore structure. Such derived physical relation between permeability and porosity from detailed pore structure information leads to a better fundamental understanding of structure-property relations in porous media.

\begin{acknowledgements}
I would like to thank Rudolf Held (Statoil) for valuable discussions and contributions to the manuscript.
\end{acknowledgements}

\bibliographystyle{spmpsci}
\bibliography{myreferences}{}

\end{document}